\documentclass[aps,prb,preprint,graphicx,amsbsy,amssymb]{revtex4}

\usepackage{graphicx}
\usepackage{graphics}% Include figure files
\usepackage{epsfig}

\begin{document}
\title{ Some Experimental Signatures to look for Time-reversal Violating
  superconductors }

\author{Tai-Kai Ng}
\address{
$^1$Department of Physics,
Hong Kong University of Science and Technology,
Clear Water Bay Road,
Kowloon, Hong Kong}

\author{C.M.Varma}
\address{University of California, Riverside, CA 92521}
\date{ \today }

\begin{abstract}

  We discuss some experimental signatures associated with the topological
  structures of unconventional superconductor order parameters of form $d_{x^2-y^2}+ix$, where $x=s,p_x \pm p_y$, or $d_{xy}$.
  Specifically, we study the  topological surface states on the $(110)$ and equivalent surfaces of such superconductors which are observable in Andreev tunneling experiments,
 as well as evaluate the magnetic flux trapped in superconducting rings of such superconductors with multiple grain-boundary Josephson
  junctions. Previous experiments are examined and several new experiments suggested.

  \end{abstract}

\maketitle

\narrowtext

Definitive experiments relying on topological differences between
different superconductors to reveal their symmetry are of interest
in themselves. The motivation for such investigations in the
Cuprates is especially strong. It has been suggested that the
pseudogap regime in the phase diagram of the Cuprates ends at a
Quantum critical point (QCP) and that the pseudogap regime
represents an elusive time-reversal violating symmetry
\cite{cmv1}. Symmetry-based proposals\cite{simon} for
experiments in the normal state using circularly polarized angle
resolved photoemission experiments have been carried out. They
show results \cite{kaminski,simon} consistent with a Time-reversal
violating phase in the normal and superconducting phases  below a
critical density.  If this is true, the superconducting order
parameter for particle density below the QCP should also be a
Time-reversal violating.

The two time-reversal violating states suggested for the
underdoped Cuprates have a current pattern sketched in fig. (1) of
Reference(\onlinecite{simon}).  State I (magnetic group in
tetragonal crystals $4/mm{\underbar{m}}$) preserves symmetry on
reflections in planes normal to the $x-y$ plane going along the
$x$ and the $y$ -axes but violates it on reflections in the
$x\pm y$ planes. It has two domains, specified by the sign of the
phase or equivalently by the direction of the current in any of
the four plaquettes in a unit-cell. State II (magnetic group
$mm{\underbar{m}}$) has current pattern which violates reflection
symmetry  in  planes going through the $x$ and the $y$ -axes. It
has four domains two due to time-reversal and two because
reflection symmetry is violated either over planes going through
the $(x+y)$ {\it or } the $(x-y)$ -axes. This state breaks
Inversion symmetry as well while State I does not. This is the
state consistent with the experiments of Kaminski et al.
\cite{kaminski}.  Considerations of these symmetries\cite{simon2,
dimatteo} and of domains is important for the design of the
experiments.

 Suppose such  states  were to undergo {\it continuous} transitions to superconductors of dominant
$d_{x^2-y^2}$ symmetry at $T_c$. Then a time-reversal violating
component would be linearly admixed which by continuity preserves
the reflection symmetry of the Time-reversal violating normal
state. For State I,  such an admixed state is uniquely of the
$d_{xy}$ variety giving a $d_{x^2-y^2} \pm i d_{xy}$
superconductor;  The $\pm$ sign signifies the two time-reversed
domains. For state II, the  admixed states possible are
either of the $p_{x+y}$ or $p_{x-y}$ variety, depending on the
domain,  giving a $d_{x^2-y^2} \pm i (p_x\pm p_y)$ superconductor.
(Due to lack of Inversion symmetry a triplet superconducting state
is linearly admixed to the singlet).  Kaur and
Agterberg\cite{kaur} have already come to these conclusions on
formal group-theoretical grounds.

  Suggestions for time-reversal breaking through admixture of the $i s$ or $id_{xy}$ type near surfaces  due to de-pairing of the principal $d_{x^2-y^2}$ order parameter and their effect on Andreev reflection experiments
have already been studied \cite{sauls}. Our principal interest here is not such surface induced time-reversal breaking which would set in at a lower temperature than the bulk $T_c$  but time-reversal breaking in the bulk, possibly existing only in cuprate superconductors for hole or electron doping less than at the quantum critical point. There have been proposals for time-reversal violation in the superconducting state alone
from different considerations \cite {laughlin, sachdev} which all
preserve center of inversion. We consider below three types
of Time-reversal violating superconducting states:
$d_{x^2-y^2}+ix$; [$x=s$, $d_{xy}$ or $p_{x\pm y}$]. The imaginary
component will be considered to be small compared to the real
part. Where the context is clear, let us call them
$d+is,d+id,d+ip$. For the first two cases the symmetry of the Andreev effect  due to bulk time-reversal breaking is the same as due to surface induced time-reversal breaking already studied \cite{sauls} but the magnitude of the effect can be quite different.

  To determine the precise form of order parameter experiments that can
   detect the presence of the small additional component  unambiguously must be designed.
 We discuss here two kinds of plausible experiments. The experimental signatures
   studied are consequences of the topological structures of the  order parameters and are robust
   to small perturbations. The first experiment  discussed is a tunnelling experiment on the $(110)$ (and
   equivalent) surfaces of the high-$T_c$ cuprates of the Andreev type. For the $x=s,d_{xy}$,  the symmetry of the Andreev effect  due to bulk time-reversal breaking is the same as due to surface induced time-reversal breaking already studied \cite{sauls} but the magnitude of the effect can be quite different.

   \subsection{surface states on $(110)$ surfaces}

    The presence of surface state with energy $\epsilon=0$ on the $(110)$ (and equivalent) surfaces of (pure)
    $d_{x^2-y^2}$-wave superconductors was first pointed out by Hu\cite{hu}. The existence of this zero energy
    surface state is a direct consequence of the topology of the superconducting order parameter. We extend Hu's
    analysis here to study surface states in superconductors with a small deviation from the pure $d_{x^2-y^2}$ symmetry.
    Following Hu's notation, we rotate our co-ordinate axis by $45^o$ on the $x-y$ plane such that the $x-$ and $y-$
    axes in the new coordinate system are the $(1\bar{1}0)$ and $(110)$ crystal axes of the cuprates, respectively.
    In this notation, the order parameter of a pure $d_{x^2-y^2}$-wave superconductor has the  property
    $\Delta(k_x,k_y,k_z)=-\Delta(-k_x,k_y,k_z)$\cite{hu}. The electronic excitations of an inhomogeneous
    superconductor are determined by the Bogoliubov-de Gennes equations,
    \begin{eqnarray}
    \label{bdg}
    \epsilon_nu_n(\vec{x}) & = & \hat{H}_ou_n(\vec{x})+\int d\vec{x}'\Delta(\vec{x},\vec{x}')v_n(\vec{x}')
    \\  \nonumber
    \epsilon_nv_n(\vec{x}) & = & -\hat{H}_ov_n(\vec{x})+\int d\vec{x}'\Delta(\vec{x},\vec{x}')u_n(\vec{x}')
    \end{eqnarray}
    where $\hat{H}_o=-{1\over2m}\nabla^2_{\vec{x}}-\mu+v(\vec{x})$, $v(\vec{x})$ is the electron crystal
    potential energy and $\Delta(\vec{x},\vec{x}')$ is a general, non-local superconductor order parameter.
    A superconductor at $x>0$ with surface at $x=0$ with thickness $d$ can be modelled by a superconductor order
    parameter of form
    \[
    \Delta(\vec{k},\vec{x})\sim\Delta_0(\vec{k})\Theta(x),  \]
    with boundary condition that the electronic wavefunctions all vanish at $x=-d$\cite{hu}.
    $\Delta_o(\vec{k}) $ is the wave-vector-dependent order-parameter of the bulk superconductor. The
    Bogoliubov-de Gennes equations for electronic states close to the Fermi surface can be solved in the WKBJ
    approximation\cite{bardeen}. In particular, surface states with
    $u(v)(\vec{x})\sim{e}^{i(k_{Fy}y+k_{Fz}z)}e^{-\gamma{x}}$ are determined by the
    following eigenvalue equation derived by Hu\cite{hu},
    \begin{equation}
    \label{eigen}
    e^{4i\epsilon{m\over{k}_{Fx}}d}=\left({\Delta_{+}\over\Delta_{-}}\right)\left({\epsilon+i\sqrt{|\Delta_{-}|^2-\epsilon^2}
    \over\epsilon-i\sqrt{|\Delta_{+}|^2-\epsilon^2}}\right),
    \end{equation}
    where $|\epsilon|<|\Delta_{+(-)}|$ is the surface state energy, $\Delta_{+}=\Delta_0(k_{Fx},k_{Fy},k_{Fz})$ and
    $\Delta_{-}=\Delta_0(-k_{Fx},k_{Fy},k_{Fz})$. $\vec{k}_F=(k_{Fx},k_{Fy},k_{Fz})$ is a wave-vector on the Fermi
    surface.

    Notice that $\epsilon$ depends on $k_{Fx},k_{Fy}$ implicitly through the momentum dependence of the
    gap function $\Delta_0(\vec{k})$. For pure d-wave superconductors, $\Delta_+=-\Delta_-$ and the eigenvalue
    equation has a particular solution $\epsilon=0$ that is independent of momentum $\vec{k}_F$ and surface
    thickness $d$. This is the zero energy mid-gap state discovered by Hu\cite{hu}. Note that other surface states
    may also exist. However they always exist in pairs with energies $\pm\epsilon$ because of particle-hole symmetry in the
    eigenvalue equation. The mid-gap states can be probed by tunnelling experiment across the (110) surface which
    measures the density of states directly. A "zero-bias conductance peak" is expected to be observed with a
    particle-hole symmetric tunnelling spectrum.

    Now consider superconductors with a small non-d-wave order parameter component. In this case the relation
    $\Delta_+=-\Delta_-$ is modified. To begin with, let $\Delta$ be real with
    $\Delta_+=\Delta+s$ and $\Delta_-=-\Delta+s$, where $s$ is independent of ${\bf k}$,  corresponding to the allowed symmetry in an orthorhombic crystal with Time-reversal preserved.
    The case of complex gap function which breaks time-reversal symmetry will soon be discussed.
    By direct inspection of Eq.\ (\ref{eigen}) we see that the $\epsilon=0$ surface states that exist for all
    momenta $k_{Fy},k_{Fz}$ for pure d-wave superconductors, survive in the presence of the small $s$-component,
    as long as $\Delta(k_{Fy},k_{Fz})>s$. Other surface states may also exist in pairs with energies $\pm\epsilon$
    because of particle-hole symmetry. Since the surface states energies must satisfy
    $|\epsilon|<\Delta(k_{Fy},k_{Fx})-s$, the main effect of mixing a small, real non-d-wave component is to reduce
    the number of available surface states that are allowed in pure d-wave superconductors. Consequently, tunnelling
    experiment cannot clearly distinguish pure d-wave superconductors from superconductors with a small, real,
    non-d-wave component.

      The situation changes if we consider superconductors with complex order parameter satisfying
    $\Delta_+=-\Delta_{-}^*$. In this case the eigenvalue equation becomes
    \begin{equation}
    \label{eigen2}
    e^{4i\epsilon{m\over{k}_{Fx}}d}=-e^{2i\phi}\left({\epsilon+i\sqrt{|\Delta_0|^2-\epsilon^2}\over
    \epsilon-i\sqrt{\Delta_{0}|^2-\epsilon^2}}\right)
    \end{equation}
    $\phi$ is defined through $\Delta_{+(-)}=|\Delta_0|e^{+(-)i\phi}$. The two signs in front of $\phi$ correspond to
    the two different domains of time-reversal or direction of flow of internal currents. The particle-hole symmetry
    that exists in pure d-wave superconductors is lost because of broken time-reversal symmetry and $\epsilon=0$
    state is no longer a solution to the equation.  Eq. (3) has the formal solution
    \begin{equation}
    \label{solution}
   2\epsilon{md\over{k}_{Fx}}-\phi=-\tan^{-1}\left({\epsilon\over\sqrt{|\Delta_{o}|^2-\epsilon^2}}\right).
    \end{equation}

      Writing $\epsilon=|\Delta_0|sin\theta$ it is easy to see that $\theta=\phi$ in the limit $d=0$
    and the surface state energy measures the imaginary part of the gap function $|\Delta_0|sin\phi$ directly.
    $\epsilon$ can be either positive or negative, depending on the sign of $\phi$. For $d\neq0$ the surface
    states do not measure the imaginary part of gap function directly and both positive and negative energy surface
    states  occur in general.

     \begin{figure}
    \leavevmode \epsfxsize9.0cm \epsfbox{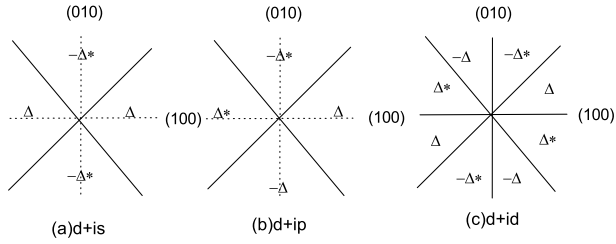}
    \caption{\label{fig:figure1} order parameter structure on the $k_x-k_y$ plane for superconductors with order
    parameter $d_{x^2-y^2}+i\delta$ for $\delta=s,p,d_{xy}$. Notice that there is another configuration for the
    $d+ip$ state where left and right region interchanges }
    \end{figure}

     Let us apply the above result to see if $x$ with different symmetries can be distinguished in
    $d_{x^2-y^2}+ix$ superconductors. In figure (1) we show the energy gap structure on the $k_x-k_y$ plane for
    superconductors with such order parameters, for $x=s,p,d_{xy}$. First we consider $d_{x^2-y^2}+is$ and
    $d_{x^2-y^2}+id_{xy}$ superconductors. Both superconductors have the property that
    $\Delta\rightarrow-\Delta^*$ upon reflection on the $(110)$ and $(1\bar{1}0)$ planes, implying absence of
    zero energy surface states and appearance of particle-hole asymmetry in the surface state spectrum for any fixed
    $k_{Fy}$. However, particle-hole asymmetry vanishes in the tunneling spectrum if the tunneling experiment samples
    all $k_{Fy}$ vectors equally, since $\Delta\rightarrow-\Delta^*$ when $k_y\rightarrow-k_y$, and the broken
    time-reversal symmetries in lower and upper $k_y$ planes are in exactly opposite domains. After adding up the
    contributions from all $k_{Fy}$ wave vectors what should be observed is a particle-hole symmetric spectrum
    with the splitting of the zero-energy bias peak into two peaks at finite energy, as pointed out in a number
    of earlier works\cite{sauls, shiba}.

     The situation is however, different for $d_{x^2-y^2}+i(p_x\pm p_y)$ superconductors. While the reflection
     symmetry of order parameter upon the $(110)$ and $(1\bar{1}0)$ planes are the same for the
     $d+is$ and $d+id$ states it is different for the $d_{x^2-y^2}+i(p_x\pm p_y)$ state.
    $\Delta\rightarrow-\Delta^*$ upon reflection on the $(110)$ plane but $\Delta\rightarrow-\Delta$ on the
    $(1\bar{1}0)$ plane or vice versa. Furthermore, the domain of broken time-reversal symmetry
    remains the same ($\Delta\rightarrow-\Delta$) when $k_y\rightarrow-k_y$. As a result, the $\epsilon=0$ surface
    states will be absent on the $(110)$ (or $(1\bar{1}0)$) surface of $d+ip$ superconductors and the corresponding
    tunnelling spectrum, after sampling over all $k_{Fy}$ vectors, will remains  particle-hole asymmetric,
    whereas $\epsilon=0$ surface states exist on the $(1\bar{1}0)$ (or $(110)$) surface with particle-hole symmetric
    tunnelling spectrum as in pure d-wave superconductors. The existence of different tunnelling spectrums on the
    $(110)$ and $(1\bar{1}0)$ surfaces with one of them being particle-hole asymmetric is a very strong indication
    of $d+ip$ superconductors. But experiments must be done in monodomain samples where $(110)$ and $(1\bar{1}0)$
    can be distinguished. There is the further problem that even in a single crystal, where for a given sign of
    time-reversal, the two different possible reflection domains $(p_x\pm p_y)$ may occur. For orthorhombic symmetry,
    this may be less likely. There is also the possibility that the boundary or tunnelling current may pin the
    orientation of the $p$ order parameter in experiment and only one of two possible surface behaviors is observable. Thus if a strong asymmetry is observed in conductance with a peak predominantly for $V>0 $ or $V<0$, a bulk $d+ip$ state may be suspected.

    For $x=is,d_{xy}$, one must also consider the possible surface induced pair-breaking \cite{sauls}. This can be
    distinguished by the temperature where it sets in, the fact that there is no obvious reason for a composition
    dependence for it and that its amplitude is expected to be smaller than the bulk effect if the thickness of the
    surface layer is small $~O$(thickness of surface layer/length scale of bound state). Otherwise experiments which
    are sensitive to the bulk (but not surface) state order parameter have to be considered. We shall examine such
    experiments in section C.

    \subsection{ Existing Andreev reflection Experiments}

    Given the above information let us look at the existing tunneling experiments to see what conclusions can be
    reached. Early Andreev reflections experiments on samples of 123-cuprate \cite {leseur} with tunneling in
    the $(110)$ and $(1\bar{1}0)$ directions have revealed symmetric peaks at zero voltage which split on applying
    a magnetic field as predicted \cite{sauls}. These confirmed that the pairing is of $d_{x^2-y^2}$  symmetry.
    Subsequently two groups \cite{dagan,grisom} have reported experiments which in zero magnetic field
    show two peaks at $\pm \delta V$ much less than the superconducting gap. The two peaks have in general different
    heights. The authors correctly conclude that a peak not at zero-voltage is an indication of time-reversal
    violating superconducting order parameter. Particularly interesting is the fact that in a given tunneling junction, the asymmetry in the amplitude of the peaks increases in annealing. Asymmetry of about a factor of 4 has been reported \cite{grisom}.  Unambiguous further experimental confirmation of the asymmetry  on monodomain samples are therefore important because one of the exact results of our analysis is that the absence of particle-hole
    symmetry in the peaks in the tunneling spectrum indicates a bulk time-reversal violating $d+ip$ state. Moreover, differences of opinion\cite{deutscher} exist about the surface composition of the samples in which the
    bound state(s) exists at zero-voltage or at finite voltage.  At this point one cannot be sure where the
   region of the peak at nonzero $V$ lies in the phase diagram  of 123-cuprates.

\subsection{Superconducting rings with multiple grain boundary Josephson junctions}

   Next  consider flux-quantization experiment on superconducting rings with multiple grain boundary Josephson
  junctions\cite{tsuei, wollman, rmp}. With suitable arrangement of the position of the grain boundaries  these
  experiments can in principle be used to distinguish between different broken time-reveral symmetry states.
  Start with a simple ring made up of two mono-crystal grains as in the Wollman et al. experiments\cite{wollman}. One grain is a simple s-wave superconductor, and the other is
  the $d+ix$-wave superconductor of interests. The arrangement of the two grains is shown in figure (3).

  \begin{figure}
  \leavevmode \epsfxsize5.0cm \epsfbox{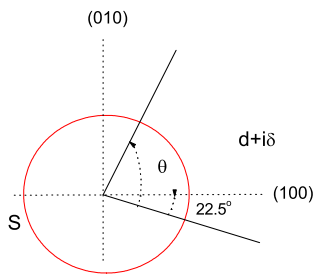}
  \caption{\label{fig:figure3} arrangement of the two superconducting grains in the two-grain ring. Notice
  that the $(100)$ axis of the $d+i\delta$-wave superconductor is $22.5^o$ off the x-direction. }
  \end{figure}

  The
  $(100)$ axis of the $d+ix$-wave superconductor is chosen to be $22.5^o$ off one of the directions of the cut. Consider the flux
  quantization condition of the ring as a function of the cutting angle $\theta$. From figure one we see that for
  $\theta<22.5^o$ the phase of the $d+ix$ superconductor remains constant over the whole region for all three
  choices of $x$. There is no mismatch in phase across the grain boundary, and consequently the flux
  trapped in the ring is integral multiple of magnetic flux quantum $\Phi_o=h/2e$, as in usual s-wave
  superconducting rings. For $22.5^o<\theta<67.5^o$, the phase of the $d_{x^2-y^2}+id_{xy}$ superconductor differs by $2\phi$ on
  the two surfaces but remains the same for the other two cases. Consequently the flux trapped in the ring is
  $(n\pm{\phi\over\pi})\Phi_o$ for the $d_{x^2-y^2}+id_{xy}$ superconductor, and remains equal to $n\Phi_o$ for the other two
  kinds of superconductors. A different regime is reached for $67.5^o<\theta<112.5^o$. In this regime, the phase
  difference between the two surfaces of the $d_{x^2-y^2}+id_{xy}$ superconductor is precisely $\pi$, whereas it is $\pi\pm2\phi$
  for $d_{x^2-y^2}+is$ superconductor, and can be either $\pi$ or $\pi\pm2\phi$ for the $d_{x^2-y^2}+ip$ superconductor, depending on
  the orientation of the $p$-order parameter. Provided that the phase $\phi$ is not too small, the different behaviors of magnetic flux trapped
  in these two regions of $\theta$-value can  distinguish $d+id$ superconductor from the other two
  types. The presence of two different possible values of trapped flux, if observable, can also
  distinguish $d+ip$ superconductor from the rest. However the magnetic flux may pin the orientation of the
  $p$-order parameter, or if only one of the two $(p_x\pm p_y)$ are present, only one type of behavior is observable. The $d+ip$ wave superconductors can be
  distinguish from the rest  if we further increase $\theta$ to $157.5^o<\theta<202.5^o$. In this
  regime, the phases on the two surfaces of the $d+ip$ superconductor differ by $2\phi$ but remain the same for the
  other two cases. The trapping of non-integral flux quantum in this region is a clear signature for $d+ip$ order
  parameter.

  Another interesting configuration that can distinguish $d+is$ and $d+id$ superconductors from the rest
  is to study Josephson effect across two $180^o$ grains in which both grains are of the cuprate superconductors
  with surfaces pinned at the $\pm(110)$ directions. This is the direction which is gapless for pure
  $d_{x^2-y^2}$-wave superconductor but are gapped for $d+is$ and $d+id$ superconductors. The existence of
  Josephson effect in this orientation provides unambiguous proof that the superconductor has a mixed order
  parameter. In this geometry, the $d+ip$ wave superconductor presents a very
  interesting possibility in principle. Since $d+ip$ wave superconductor is gapless along one of the
  $(110)$ or $(1\bar{1}0)$ directions and gapped in the other, Josephson effect with phase shift near $\pi/2 \pm \phi$ is to be expected in one
  orientation and no Josephson effect in the other.  There are probably experimental limitations for such
  observations which we are not experts on, as well as the possibility that the orientation of the $p$ order
  parameter may be pinned by the boundary.

  \begin{figure}
  \leavevmode \epsfxsize6.0cm \epsfbox{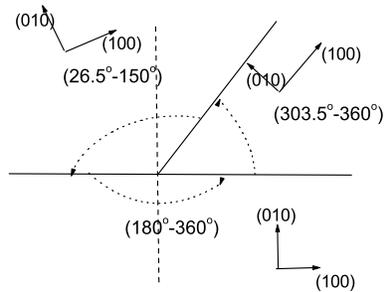}
  \caption{\label{fig:figure4} arrangement of the three superconducting grains in the experiment by Tsuei
  {\em et.al.} }
  \end{figure}

    Finally, let us study the situation of a 3-junction ring in the celebrated geometry first fabricated by Tsuei
  {\em et.al.}\cite{tsuei}. The ring is made by the superconductor of interests only, but with different crystal
  orientations in the three grains. On the $k_x-k_y$ plane the three grains cover regions specified by angles
  $180^o-360^o$, $303.5^o-360^o$, and $26.5^o-150^o$, ($0^o$ is the $(110)$ axis) respectively (see figure three).
  For pure $d$-wave superconductor there is a phase difference of $\pi$ on the two surfaces of the second grain
  but the phases are the same on the two surfaces of other two grains. Consequently the flux trapped in the ring is
  precisely $(n+{1\over2})\Phi_o$.

    Repeating the analysis for the three different kinds of time-reversal violating superconductors of interest here, we find that the
  flux trapped in the corresponding rings of $d+is$ superconductor is given by
  $F_{s}=(n+{1\over2}\pm{\phi\over\pi})\Phi_o$, whereas the flux trap in the $d+ip$ superconductor can be
  either $F_p=(n+{1\over2})\Phi_o$ or $(n+{1\over2}\pm{\phi\over\pi})\Phi_o$, depending on the orientation of the
  $p$-order parameter. The flux trap in the $d+id$ superconductor is the same as $d+is$ superconductor,
  $F_{d}=(n+{1\over2}\pm{\phi\over\pi})\Phi_o$. However, there is some ambiguity because of the node in $id_{xy}$
  order parameter along $x$ and $y$ axes. Note that with different cutting angles the value of the trapped flux
  will change. For example, if the three grains cover regions specified by angles $180^o-360^o$, $280^o-350^o$,
  and $60^o-170^o$, then the flux trapped in a pure d-wave superconductor ring will be $n\Phi_o$, whereas the
  $d+ip$ superconductor will trap flux $F_p=(n\pm{\phi\over\pi})\Phi_o$, independent of the orientation of the
  $p$-order parameter.

  Recently, Tsuei {\em et.al.} have reported results of  experiments on superconducting rings with essentially
  their original 3-junction ring geometry for three different cuprate systems\cite{tsuei2} for various dopings. They observed that the
  trapped flux is always equal to $(n+{1\over2})\Phi_o$ within experimental accuracy of a few percent. Based on this observation,
  they conclude that the mixing of $d_{x^2-y^2}+is$ and $d_{x^2-y^2}+id_{xy}$ states, if exists, must be small in
  a large range of dopping in high-$T_c$ cuprates. While we agree with their conclusion on the $d_{x^2-y^2}+is$ and
  $d_{x^2-y^2}+id_{xy}$ states, our analysis shows that the $d_{x^2-y^2}+i(p_x\pm p_y)$ state is
  not ruled out by their results. We therefore suggest experiments with different cutting angles as discussed above. These may also be more precise  since the dominant flux due to the dominant d-wave symmetry is nulled out in such experiments and only the corrections are visible.

{\it Acknowledgements}
 C.M. Varma wishes to thank Guy Deutscher,  Brigitte Leridon and Jerome Lesueur for interesting him in this investigation and for various useful comments. T.K. Ng acknowledge support of HKRGC through Grant 602803.


\references
\bibitem{cmv1} C.M. Varma, \prb {\bf 55}, 14554 (1997); \prl {\bf 83},3538 (1999).
\bibitem{simon} M.E. Simon and C.M. Varma, \prl {\bf 83}, 247003-1 (2002); C.M.Varma, Phys. Rev. B {\bf 61}, R-3804 (2000).
\bibitem{kaminski} A. Kaminski et al., Nature {\bf 416}, 610 (2002).
\bibitem{simon2} M.E. Simon and C.M. Varma , \prb {\bf 67} 054511 (2003)
\bibitem{dimatteo} S.Di Matteo and C.M. Varma, \prb {\bf 67} 134502 (2003).
\bibitem{kaur} R.P. Kaur and D.F. Agterberg, \prb {\bf 68} 100506 (2003); cond-mat/0306677.
\bibitem{laughlin}R. Laughlin, \prl {\bf 80}, 5188 (1998).
\bibitem{sachdev} M. Vojta {\em et al.}, \prb {\bf 62},6721 (2000).
\bibitem{sauls} M. Fogelstrom, D. Rainer and J.A. Sauls, \prl {\bf 79}, 281 (1997).
\bibitem{hu} C.-R. Hu, \prl {\bf 72}, 1526 (1994).
\bibitem{bardeen} J. Bardeen {\em et.al.}, \prb {\bf 187}, 556 (1969).
\bibitem{shiba} M. Matsumoto and H. Shiba, J. Phys. Soc. Jpn. {\bf 64}, 4867 (1995).
\bibitem{leseur} J. Leseur et al., Physica, {bf 191C}, 325 (1992); M. Aprili, E.Badica and L.H.Greene, \prl {\bf 83},
  4630 (1999); R. Krupke and G. Deutscher, \prl {\bf 83}, 4634 (1999); M. Covington {\em et. al.}, \prl {\bf 79},
  277 (1997).
\bibitem{dagan} Y. Dagan and G. Deutscher, \prl {\bf 87}, 17700-1 (2001).
\bibitem{grisom} X. Grisom, Thesis, Ecole Polytechnic (Paris), November (2000).
\bibitem{deutscher} Private communications with Guy Deutscher, Laura Greene, Brigitte Leridon and
Jerome Leseur.
\bibitem{tsuei} C.C. Tsuei {\em et. al.}, \prl {\bf 73}, 503 (1994).
\bibitem{wollman}  D.A. Wollman {\em et al.}, \prl {\bf 74}, 697 (1995).
\bibitem{rmp} C.C. Tsuei and J.R. Kirtley, Rev. Mod Phys.  {\bf 72}, 969 (2000).
\bibitem{tsuei2} C.C. Tsuei {\em et.al.}, preprint (cond-mat/0402655).
\end{document}